\documentclass[aps,prl,reprint,superscriptaddress,footinbib,10pt,showpacs,floatfix,longbibliography]{revtex4-2}

\pdfoutput=1

\usepackage[T1]{fontenc} 
\usepackage[english]{babel} 
\usepackage[utf8]{inputenc}

\usepackage[table,dvipsnames]{xcolor}
\usepackage[psdextra]{hyperref}
\usepackage{amsmath,amssymb,amsfonts,mathtools,amsthm}
\usepackage{graphicx}
\usepackage{bbm}

\usepackage{environ}
\usepackage{enumitem}
\usepackage[normalem]{ulem}

\mathtoolsset{showonlyrefs}

\newtheorem{theorem}{Theorem}
\newtheorem*{theorem*}{Theorem}

\theoremstyle{definition}

\DeclareMathOperator{\tr}{tr}

\newcommand{\fro}{\mathrm{F}}

\DeclareMathOperator{\U}{U}

\newcommand{\haar}{\mathrm{H}}

\NewDocumentCommand\Cl{mm}{
    \ensuremath{\mathrm{Cl}_{#1}\IfNoValueTF{#2}{}{(#2)}}%
}
\NewDocumentCommand\HW{mm}{
    \ensuremath{\mathrm{HW}_{#1}\IfNoValueTF{#2}{}{(#2)}}%
}

\newcommand{\EE}{\mathbb{E}}

\newcommand{\C}{\mathbb{C}}
\newcommand{\F}{\mathbb{F}}

\newcommand{\one}{\mathbbm{1}}

\newcommand{\argdot}{{\,\cdot\,}}

\newcommand{\myleft}{\mathopen{}\mathclose\bgroup\left}
\newcommand{\myright}{\aftergroup\egroup\right}

\DeclarePairedDelimiterX{\abs}[1]{\lvert}{\rvert}{%
  \ifblank{#1}{\,\cdot\,}{#1}
}   

\DeclarePairedDelimiterX\norm[1]\lVert\rVert{%
  \ifblank{#1}{\,\cdot\,}{#1}
}   

\DeclarePairedDelimiterX{\iiiNorm}[1]{\lvert}{\rvert}{%
  \delimsize\lvert\delimsize\lvert#1\delimsize\rvert\delimsize\rvert%
}

\DeclarePairedDelimiterXPP\snorm[1]{}\lVert\rVert{_\infty}{\ifblank{#1}{\,\cdot\,}{#1}}   

\DeclarePairedDelimiterXPP\twonorm[1]{}\lVert\rVert{_2}{\ifblank{#1}{\,\cdot\,}{#1}}   

\DeclarePairedDelimiterXPP\trnorm[1]{}\lVert\rVert{_1}{\ifblank{#1}{\,\cdot\,}{#1}}   

\DeclarePairedDelimiterXPP\fnorm[1]{}\lVert\rVert{_{\fro}}{\ifblank{#1}{\,\cdot\,}{#1}}   

\DeclarePairedDelimiterXPP\dnorm[1]{}\lVert\rVert{_\diamond}{\ifblank{#1}{\,\cdot\,}{#1}}   

\DeclarePairedDelimiterXPP\cbnorm[1]{}\lVert\rVert{_\mathrm{cb}}{\ifblank{#1}{\,\cdot\,}{#1}}   
\DeclarePairedDelimiterXPP\onenorm[1]{}\lVert\rVert{_{1\rightarrow 1}}{\ifblank{#1}{\,\cdot\,}{#1}}   
\DeclarePairedDelimiterXPP\ddnorm[1]{}\lVert\rVert{_{\diamond\rightarrow \diamond}}{\ifblank{#1}{\,\cdot\,}{#1}}   
\DeclarePairedDelimiterXPP\ssnorm[1]{}\lVert\rVert{_{\infty\rightarrow\infty}}{\ifblank{#1}{\,\cdot\,}{#1}}   

\DeclarePairedDelimiterX\Set[1]\{\}{%
  
  #1
}

\DeclarePairedDelimiterX\innerp[2]{\langle}{\rangle}{%
  \ifblank{#1}{\,\cdot\,}{#1} , \ifblank{#2}{\,\cdot\,}{#2}%
}

\DeclarePairedDelimiter{\ket}{\vert}{\rangle}

\DeclarePairedDelimiterX\braket[2]{\langle}{\rangle}%
  {#1\kern0.15ex\delimsize\vert\kern0.15ex\mathopen{}#2}

\DeclarePairedDelimiterX\ketbra[2]{\vert}{\vert}%
  {#1\kern0.15ex\delimsize\rangle\delimsize\langle\kern0.15ex\mathopen{}#2}

\DeclarePairedDelimiterX\sandwich[3]{\langle}{\rangle}%
  {#1\,\delimsize\vert\kern0.15ex\mathopen{}#2\kern0.15ex\delimsize\vert\kern0.15ex\mathopen{}#3}

\newcommand{\out}[1]{\ketbra{#1}{#1}}

\DeclarePairedDelimiterX\obraket[2]{(}{)}%
  {#1\kern0.15ex\delimsize\vert\kern0.15ex\mathopen{}#2}

\DeclarePairedDelimiterX\oketbra[2]{\vert}{\vert}%
  {#1\kern0.15ex\delimsize)\delimsize(\kern0.15ex\mathopen{}#2}

\DeclarePairedDelimiterX\osandwich[3]{(}{)}%
  {#1\,\delimsize\vert\kern0.15ex\mathopen{}#2\kern0.15ex\delimsize\vert\kern0.15ex\mathopen{}#3}

\newcommand{\Sym}{\vee}
\newcommand{\Alt}{\wedge}
\newcommand{\Psym}{P_{\Sym}}
\newcommand{\Dsym}{D_{\Sym}}
\newcommand{\Palt}{P_{\Alt}}
\newcommand{\Dalt}{D_{\Alt}}

\newcommand{\MO}{\mathsf{M}}
\newcommand{\mo}{\mathsf{m}}

\begin{document}

\newcommand{\thetitle}{Anticoncentration is (almost) all you need}
\title{\thetitle}
\author{Markus Heinrich}
\email{markus.heinrich@uni-koeln.de}
\affiliation{Institute for Theoretical Physics, University of Cologne, Germany}
\author{Jonas Haferkamp}
\affiliation{School of Engineering and Applied Science, Harvard University, USA}
\affiliation{Department of Mathematics, Saarland University, Germany}
\affiliation{Department of Computer Science, Ruhr-University Bochum, Germany}
\author{Ingo Roth}
\affiliation{Quantum Research Center, Technology Innovation Institute, Abu Dhabi, United Arab Emirates}
\author{Jonas Helsen}
\affiliation{QuSoft and CWI, Amsterdam, Netherlands}

\begin{abstract}
Until very recently, it was generally believed that the (approximate) 2-design property is strictly stronger than anticoncentration of random quantum circuits, mainly because it was shown that the latter anticoncentrate in logarithmic depth, while the former generally need linear depth circuits.
This belief was disproven by recent results 
which show that so-called \emph{relative-error} approximate unitary designs can, in fact, be generated in logarithmic depth, implying anticoncentration.
Their result does however not apply to ordinary local random circuits, a gap which we close in this letter, at least for 2-designs.
More precisely, we show that anticoncentration of local random quantum circuits already implies that they form relative-error approximate state 2-designs, making them equivalent properties for these ensembles.
Our result holds more generally for any random circuit which is invariant under local (single-qubit) unitaries, independent of the architecture.
\end{abstract}

\maketitle

Random quantum circuits are ubiquitous in quantum information theory, touching a wide range of topics from randomized benchmarking \cite{emerson_scalable_2005,knill_randomized_2008,magesan_characterizing_2012,helsen_general_2022,heinrich_randomized_2023} to black hole~\cite{brown_second_2018,akers_holographic_2024} and many-body physics \cite{fisher_random_2023}. 
Consequently, significant effort has been devoted to studying the mixing properties of random quantum circuits.

A particular focus of the community concerns the convergence of random quantum circuits to approximate (unitary or state) $k$-designs.
These are probability distributions on the unitary group or state space that appear uniformly random given access to, at most, $k$ copies. 
Early works rigorously proved the convergence of random quantum circuits to approximate $2$-designs in depth $O(n)$ on $n$ qubits \cite{harrow_random_2009,dankert_exact_2009}.
The convergence was subsequently tied to spectral properties of the $k$-copy twirling channel \cite{brown_convergence_2010,brown2012scrambling}.
A long line of works \cite{brandao2016local,harrow_approximate_2023,haferkamp_improved_2021,haferkamp_random_2022} eventually resulted in near tight bounds on the $k$-dependence~\cite{chen2024incompressibility}.
At the same time, the linear dependence on the system size was widely considered optimal.
This, however, changed very recently, when it was shown that random quantum circuits may already look Haar-random to forward-in-time experiments at logarithmic depth \cite{schuster_random_2025,laracuente_approximate_2025}.

In hindsight, the equally rapid onset of \emph{anticoncentration}~\cite{dalzell_random_2022,barak_spoofing_2021} could have been seen as a first hint for the logarithmic convergence of second moments. 
Anticoncentration 
refers, loosely speaking, to the property that the outcome distribution of randomly sampled circuits does not have too many probabilities that are ``too small'' compared to the uniform distribution (e.g.~only a small fraction of the outcome probabilities can be exactly zero).
This property is a key ingredient in complexity-theoretic arguments for sampling-based quantum advantage (see~Ref.~\cite{hangleiter_computational_2023} for a recent review). 
More technically, it allows to reduce the hardness of approximating outcome probabilities of quantum circuits in additive error to approximating them in relative error, providing the basis to apply further techniques for proving hardness-of-estimation. 

The logarithmic-depth designs in Refs.~\cite{schuster_random_2025,laracuente_approximate_2025}, as well as all of the more recent follow-up works~\cite{cui2025random,cui2025unitary,schuster2025strong,mao2025random}, are based on specially structured circuits.
In contrast, essentially no progress has been made on unstructured local circuits such as the infamous brickwork circuits.
From the latter, the structured circuits in Refs.~\cite{schuster_random_2025,laracuente_approximate_2025} can be obtained by deleting entangling gates over certain cuts. This raises the question whether this deletion is actually advantageous for the convergence to designs, or whether it is simply a remnant of available proof techniques.
Intuitively, the removal of random gates should slow down the mixing time, but contrary effects are known to occur in Markov chains (cf.~Ref.~\cite[{Sec.~1.5}]{harrow_approximate_2023}).
A similar behavior was recently observed in quantum circuits involving Haar-random unitaries that act on an extensive number of qubits~\cite{belkin_absence_2025}.

In this letter, we provide the first evidence that standard random quantum circuits indeed converge as fast as the coarse-grained circuits in Refs.~\cite{schuster_random_2025,laracuente_approximate_2025}, demonstrating that this extra structure is not necessary and that the result of Ref.~\cite{belkin_absence_2025} does not qualitatively apply.
Our proof is remarkably short and reduces the relative-error state $2$-design property to the onset of \emph{anticoncentration}.
Consequently, standard brickwork random quantum circuits generate relative-error state $2$-designs in logarithmic depth.
Interestingly, the inverse implication --being a relative-error $2$-design implies anticoncentration-- is always true \cite[Theorem 5]{hangleiter_anticoncentration_2018} (see also the discussion below Eq.~\eqref{eq:relative-error-psd-moments}).
Hence, \emph{anticoncentration is all you need} for the here considered state ensembles.
Finally, we also briefly discuss the case of unitary designs and why anticoncentration may generally not be enough there.
These intricacies are illustrated in parallel work \cite{belkin_absence_2025} which also gives numerical evidence that anticoncentration implies the unitary 2-design property in 1D brickwork circuits.

Beyond the immediate relevance for understanding the convergence behavior of random quantum circuits, our result also shows that notoriously hard-to-study designs can be reduced to a single and much more accessible \emph{universal property}:  anticoncentration, or more concretely, the so-called \emph{collision probability}.
The latter is also known as \emph{inverse participation ratio} in the many-body literature and, for most circuit ensembles used in practice, is equal to the \emph{frame potential} (up to a factor of two in the circuit depth).
The latter concepts are central to a rich literature in many-body physics and quantum information \cite{gross2007evenly,roberts_chaos_2017,cotler_chaos_2017,hunter-jones_unitary_2019,haferkamp2021emergent,fisher_random_2023,leone_non-clifford_2025,sauliere_universality_2025,lami_anticoncentration_2025,magni_anticoncentration_2025,mok_optimal_2025,lami_quantum_2025,dowling_free_2025,magni_anticoncentration_2026,feng_thermalization_2025}, mostly because of their computational accessibility despite a lack of operational meaning.
While it was previously known that relative errors in the frame potential bound the distinguishability from Haar-random states, our results go beyond that and precisely equate them with relative-error designs.
This gives a strong retrospective justification for these measures and greatly extends the implications of prior results.
Our result also explains the predictive power of the anticoncentration property, for instance for efficiently simulating samples from the outcome distributions of noisy random quantum circuits~\cite{aharonov2023polynomial}.\\

\paragraph*{Preliminaries.} We consider a system of $n$ qudits of local dimension $q$. 
Given a probability measure $\nu$ on $\U(q^n)$, the (average) \emph{collision probability} $\mathbb{E}_{U\sim \nu} \sum_{x\in [q]^n} p_x(U\ket{0})^2$ is the probability that a given outcome $x$ is observed twice (a collision) upon measuring the state $U\ket{0}$ in the computational basis. 
We want to assume that this quantity does not depend on $x$ ($\nu$ has the \emph{hiding property}), e.g.~because $\nu$ is invariant under $X$ gates.
Then, the collision probability is, up to a dimensional factor, equal to:
\begin{equation}
\label{eq:collision-probability}
 Z_\nu 
 = \EE_{U\sim\nu} | \sandwich{0}{U}{0} |^4 \,.
\end{equation}
In a slight abuse of language, we will also call $Z_\nu$ the collision probability.
It
is minimal for circuits producing uniform outcome distributions (e.g.~a layer of Hadamards), $Z_\mathrm{uni} = q^{-2n}$, while for Haar-random unitaries, we have $Z_\haar = 2 q^{-n} (q^n+1)^{-1}$.
We say that $\nu$ (strongly) \emph{anticoncentrates} if $Z_\nu \leq \alpha q^{-2n}$ for some $\alpha \geq 1$.
This implies that the outcome distributions of such circuits anticoncentrate in the original sense, i.e.~a concentration of probability cannot happen for at least a constant fraction of instances \cite{hangleiter_anticoncentration_2018}.
Although being strictly weaker, anticoncentration is often equated with the convergence of the collision probability in the random circuits literature and we follow this convention in this paper.
While, generally speaking, any constant $\alpha \geq 1$ will do, we here consider the case $Z_\nu \leq ( 1 + \varepsilon) Z_\haar$ with $\varepsilon \in [0,1)$.
It is expected that any constant $\varepsilon$ can be achieved at the cost of a constant overhead in the total number of gates \cite{dalzell_random_2022,barak_spoofing_2021}.

A probability measure $\nu$ on $\U(q^n)$ is called a \emph{relative-error $\varepsilon$-approximate unitary $k$-design} (or, for short, \emph{relative-error design}) if the $k$-fold twirling channel $\MO_{k,\nu} := \EE_{U\sim\nu} U^{\otimes k}(\argdot)U^{\dagger\otimes k}$ fulfills the operator inequalities
\begin{equation}
\label{eq:def-relative-design}
 (1-\varepsilon) \MO_{k,\haar} \leq_\mathrm{CP} \MO_{k,\nu} \leq_\mathrm{CP} (1+\varepsilon) \MO_{k,\haar} \,,
\end{equation}
where $\mathcal{A} \leq_\mathrm{CP} \mathcal{B}$ iff $\mathcal{B}-\mathcal{A}$ is completely positive (CP), and the index $\haar$ refers the integration w.r.t.~the Haar measure $\mu_\haar$ on $\U(q^n)$.
To see that $\varepsilon$ is indeed a relative error, consider positive-semidefinite (psd) operators $A,B\geq0$.
Then, the definition \eqref{eq:def-relative-design} implies that
\begin{equation}
\label{eq:relative-error-psd-moments}
 \left| \frac{\tr(A \, \MO_{k,\nu}(B) ) - \tr(A \, \MO_{k,\haar}(B) ) }{\tr(A \, \MO_{k,\haar}(B))} \right| \leq \varepsilon \,,
\end{equation}
i.e.~all \emph{psd Haar moments} are approximated within relative error $\varepsilon$.
Moreover, we say that the generated state ensemble $\{U\ket{0}\}_{U\sim\nu}$ is a \emph{relative-error $\varepsilon$-approximate state $k$-design} if Eq.~\eqref{eq:relative-error-psd-moments} holds for any $A\geq 0$ and $B=\out{0}^{\otimes k}$ (here $0\equiv 0^n$ denotes the all-zero state).
Setting $k=2$ and $A = B = \out{0}^{\otimes 2}$ in Eq.~\eqref{eq:relative-error-psd-moments}, we recover the well-known fact that relative-error (state or unitary) 2-designs with error $\varepsilon$ anticoncentrate with $\alpha = 2(1+\varepsilon)$.

In this paper, we will focus on probability measures $\nu$ that are generated by \emph{local random quantum circuits} (local RQCs). These are circuits that are composed of Haar-random 2-local unitaries arranged in a prespecified manner.
These form a significant subclass for design constructions, with additional applications in many-body physics \cite{nahum_operator_2018,zhou_emergent_2019,hunter-jones_unitary_2019,dalzell_random_2022}. 
Using a mapping to a statistical mechanics model \cite{nahum_operator_2018,hunter-jones_unitary_2019,zhou_emergent_2019}, it was shown in Refs.~\cite{dalzell_random_2022,barak_spoofing_2021} that local random quantum circuits anticoncentrate already at logarithmic depth.

We will further assume that the RQC is invariant under local, single-qudit (Clifford) unitaries (LU)--this is simply to avoid technicalities on the support of single instances of the RQCs.
This is true for many ensembles studied in the literature \cite{brandao2016local,suzuki_more_2024,schuster_random_2025,haah2025efficient} and can be straightforwardly imposed by a layer of single-qudit gates at the start of the circuit.
We note that our results generalize to more structured RQCs as well, where the local gates are drawn from a gate set instead of Haar-randomly \cite{suzuki_more_2024}.\\

\paragraph*{State designs.}
We show that if a local RQC anticoncentrates, it also forms a relative-error state 2-design: 

\begin{theorem}
\label{thm:main}
Let $\nu$ be the probability measure of a local RQC on $n$ qudits and suppose it anticoncentrates in the sense that $Z_\nu \leq ( 1 + \varepsilon) Z_\haar$.
Then, $\nu$ generates a relative-error $\varepsilon'$-approximate state 2-design, where $\varepsilon' = 2 \, \frac{q^n+1}{q^n-q} \frac{\varepsilon}{1-q^{-1}} \approx 4 \varepsilon$.
\end{theorem}

The theorem immediately implies that local RQCs in a 1D nearest-neighbor or all-to-all architecture form relative-error state 2-designs in logarithmic depth.

The argument is 
simple, centered around a single application of H\"older's inequality.
For the sake of notation we set $\MO_{\nu} \equiv \MO_{2,\nu}$ in the following.

\begin{proof}
We first establish some facts about the collision probabilities $Z_\nu$ and $Z_\haar$.
Note that the LU invariance implies that $\mo_\nu := \MO_\nu(\out{0}^{\otimes 2})$ commutes with $(U_1\otimes\dots\otimes U_n)^{\otimes 2}$ for $U_i\in\U(q)$.
Thus, applying Schur-Weyl duality locally on every qudit, we can expand $\mo_\nu$ in the local permutation basis $\{\one, F\}^{\otimes n}$, where $F$ is the flip operator permuting two tensor copies of $\C^q$:
\begin{equation}
\label{eq:permutation-expansion}
 \mo_\nu 
 =
 \sum_{x\in\F_2^n} m_x F_x\,,
\end{equation}
with $F_x := \bigotimes_{i=1}^n F^{x_i}$ and $\F_2$ is the binary field.
Denoting the canonical 
dual basis by $\{\hat F_x\}$, we can write the coefficients as $m_x = \tr(\hat F_x \mo_\nu)$.
As any $F_x$ acts trivially on $\ket{0}^{\otimes 2}$, we have the relation $Z_\nu = \tr(\ketbra 0 0^{\otimes 2} \mo_\nu) =  \sum_{x\in\F_2^n} m_x$. 
Note that $\mo_\haar:=\MO_\haar(\out{0}^{\otimes 2})$ only features contributions from $x=0$ and $x=1$, the all-zero and all-one bitstrings.
We can, thus, expand $Z_\haar = \tr( \hat F_0 \mo_\haar) + \tr( \hat F_1 \mo_\haar)$ and use that $\MO_\haar = \MO_\haar \MO_\nu$ by the invariance of the Haar measure.
Then, we find
\begin{align}
 Z_\haar
 &=
 \tr\left( (\hat F_0 + \hat F_1) \MO_\haar( \mo_\nu )  \right) \\
 &=
 \sum_{x\in\F_2^n}
 \tr\left( (\hat F_0 + \hat F_1) \MO_\haar( F_x )  \right) \tr(\hat F_x \mo_\nu) \\
 &=
 m_0 + m_1 + \sum_{x\notin \{0,1\}} \alpha_{|x|} m_x \,, \label{eq:Zhaar-expansion}
\end{align}
where the last line follows from writing out $\MO_\haar$ in the local permutation basis (see the Supplemental Material for details).
Here, $\alpha_{|x|} := \frac{q^{|x|}+q^{n-|x|}}{q^n+1} $ and $|x|$ is the Hamming weight of the binary vector $x$.
Using that the maximum of $\alpha$ over $x\neq0,1$ is attained at $|x|=1$ and $Z_\nu \leq (1+\varepsilon) Z_\haar$, we have
\begin{align}
 \sum_{x\notin \{0,1\} } m_x
 &=
 Z_\nu - m_0 - m_1 \\
 &\leq
 (1+\varepsilon) Z_\haar - Z_\haar + \alpha_1 \sum_{x\notin \{0,1\} } m_x \,. 
\end{align}
With $\alpha_1 < 1$ we then find the following bound
\begin{equation}
\label{eq:sum-wx-bound}
 \sum_{x\notin \{0,1\} } m_x 
 \leq
 \frac{\varepsilon}{1-\alpha_1} Z_\haar
 =: \frac{\varepsilon'}{2} Z_\haar \,.
\end{equation}
Explicitly, we have $\varepsilon' = 2 \, \frac{q^n+1}{q^n-q} \frac{\varepsilon}{1-q^{-1}}$ where $\frac{q^n+1}{q^n-q}$ is quickly converging to $1$ from above and $\frac{1}{1-q^{-1}}\leq 2$.
Thus, for sufficiently large $n$ (say $n \approx 10$), $\varepsilon' \approx 4 \varepsilon$.

Finally, we bound the relative error of $\tr(A \mo_\nu)$ for any psd operator $A$.
For local RQCs, it was shown in Ref.~\cite{dalzell_random_2022} that the $m_x = \tr(\hat F_x \mo_\nu)$ can be computed using a statistical mechanics model and are non-negative numbers depending on the architecture and depth of the circuit. 
With this and the expansions \eqref{eq:permutation-expansion} and \eqref{eq:Zhaar-expansion}, a simple application of Hölder's inequality yields
\begin{align}
 &|\tr(A \mo_\nu) - \tr(A \mo_\haar) | \\
 &\quad=
  \Big| \sum_{x\notin \{0,1\} } \Big( \tr(A F_x) - \tr(A) \alpha_{|x|} \Big) m_x \Big| \\
 &\quad\leq
  2 \tr(A) \sum_{x\notin \{0,1\} } m_x \\
 &\quad\leq
  \varepsilon' \tr(A) Z_\haar = \varepsilon' \tr(A \mo_\haar) \,,
\end{align}
where we used Eq.~\eqref{eq:sum-wx-bound} in the last line and the fact that $\tr(A\mo_\haar) = \tr(A) Z_\haar$ since $\mo_\haar$ is proportional to the projector onto the global symmetric subpace.
This completes the argument.
\end{proof}

We think that the factor in front of $\varepsilon$ can be improved to $1$, as we can, rather trivially, upper bound $\tr(A \mo_\nu)$ using Hölder's inequality as follows:
\begin{equation}
 \tr(A \mo_\nu)
 =
 \sum_x \tr(A F_x) m_x 
 \leq
 \tr(A) Z_\nu
 \leq
 (1+\varepsilon) \tr(A \mo_\haar) \,.
\end{equation}
Obtaining the lower bound however requires a more careful analysis.

We remark that our proof works for any measure $\nu$ for which the expansion \eqref{eq:permutation-expansion} has non-negative coefficients $m_x$.

This is, in particular, the case if the local moment operator associated with the distribution of 2-local gates has non-negative matrix coefficients.
This is true
for Haar-random $2$-local gates as we assumed in Thm.~\ref{thm:main}, but also for more structured circuits \cite{suzuki_more_2024}.\\

\paragraph*{Unitary designs.}

Unfortunately, a straightforward extension of the ideas in the last section to the unitary 2-design case is not possible. 
In fact, repeating the above steps leads to an exponential blow-up in the relation between the collision probability and the design error.
To understand why such an argument might be difficult, we will briefly discuss possible strategies in the following.
To do so, it will be convenient to change the basis to the one given by the mutually orthogonal local projectors $P_a := \otimes_{i=1}^n P_{a_i}$, $a\in\F_2^n$ with $P_0 = \frac12(\one + F)$ and $P_1 = \frac12(\one - F)$.
We denote their rank as $D_a = 2^{-n} q^{n} (q-1)^{|a|} (q+1)^{n-|a|}$.

Because of the local $\U(q)$ invariance and $P_a\geq 0$, it is sufficient to verify Eq.~\eqref{eq:relative-error-psd-moments} on the \emph{local projector basis} (see the Supplemental Material for details).
We thus aim to bound the expression
\begin{equation}
   \varepsilon \leq \max_{a,b} \frac{|\tr(P_a (\MO_\nu - \MO_\haar)(P_b))|}{\tr(P_a \MO_\haar(P_b))} \,.
\end{equation}
This already reduces the problem to bounding a finite number of moments. However we found no easy way to do so even for $1$D brickwork circuits.
Let us now make the simplifying assumption that $\MO_\nu$ is a psd 
superoperator.
This is true, for instance, for random circuits composed of a single Haar-random local gate per layer, for 1D brickwork circuits with an odd number of layers, or more generally for ensembles that are invariant under inverses (if one is willing to double the depth of the circuit).
Under this assumption, \textcite{belkin_absence_2025} show that it is sufficient to probe only the `diagonal' elements in the local projector basis, i.e.~the relative error is given by 
\begin{align}
    \varepsilon
    &=
    \max_{a,b}\frac{|\tr(P_a (\MO_\nu - \MO_\haar)(P_b))|}{\tr(P_a \MO_\haar(P_b))} \\
    &=
    \max_a \frac{\tr(P_a \MO_\nu(P_a))}{\tr(P_a \MO_\haar(P_a))} - 1 \\
    &=
    q^n \max_a \left(q^n + (-1)^{|a|}\right) \frac{\tr(P_a \MO_\nu(P_a))}{2 D_a^2} - 1 \,. \label{eq:relative-error}
\end{align}
Here, we used that $\MO_\nu - \MO_\haar$ is psd since it has the same spectrum as $\MO_\nu$, except that two `1' eigenvalues are set to zero. 
Moreover,  $\tr(P_a \MO_\haar(P_a)) = 2 D_a^2 q^{-n}/(q^n + (-1)^{|a|})$. 

We remark that the term with $a=0$ exactly corresponds to the relative anticoncentration error $Z_\nu/Z_\haar - 1$.
A priori, it is not clear how this error should bound the maximum over all $a$.
Indeed, while $\tr(P_a \MO_\nu (P_a))/D_a$ attains its maximum at $a = 0$ (see Supplemental Material), the remaining factor in Eq.~\eqref{eq:relative-error} is maximized at $a = 1$ for large $n$. 
This leads to a competition between these two terms and, thus, to a complex behavior.
Numerical studies \cite{belkin_absence_2025} show that the behaviors of the anticoncentration error and the relative design error $\varepsilon$ can be very different and generally depends highly on the concrete random circuit ensemble and the connectivity.
This indicates that anticoncentation and relative unitary designs might be less related than one might hope from our results on state designs.\\

\paragraph*{Discussion and outlook.}
In this letter, we show that anticoncentration of local random quantum circuits implies the approximate $2$-design property for the generated states.
We provide the first evidence that the deletion of local gates in Ref.~\cite{schuster_random_2025} does not provide an advantage over ordinary brickwork circuits in generating designs and pseudorandom unitaries.
Complementary evidence for the fast convergence of unstructured random quantum circuits was recently obtained in Ref.~\cite{laracuente_approximate_2025}, which shows that the structure of the coarse-grained circuits in Refs.~\cite{schuster_random_2025,laracuente_approximate_2025} does not change the relative entropy decay too much. 
This implies additive-error designs (in diamond norm) in depth $\mathrm{polylog}(n)$, even for higher $k$.

The fast convergence to approximate state designs implies many intuitive properties of states generated by shallow random quantum circuits, whose proof remained elusive until now.
First, 2-designs are well known to generate near-maximal entanglement across any bipartite cut.
Consequently, our result shows that random quantum circuits of depth $d$ in a brickwork layout generate as much entanglement as possible with circuits of depth $d$ up to log-factors.
Moreover, the variance of expectation values $\mathrm{Tr}[O \psi]$ of bounded observables is a second moment quantity and, therefore, the state 2-design property implies concentration results (see e.g. Ref.~\cite{haferkamp2021emergent}). 
Another consequence of the second moments converging is that it implies equilibration under the time evolution of many natural Hamiltonians~\cite{reimann2008foundation}. 
Finally, the $2$-design property implies superpolynomial sampling complexity for property testing of random brickwork circuits of super-logarithmic depth: In Refs.~\cite{huang2022quantum,chen2022exponential} it was proven that exponentially many copies are required to distinguish an exact state 2-design from the maximally mixed state using unentangled measurements.

Interestingly, anticoncentration is not necessarily universal for local random circuits over more restricted gate sets.
For instance, if we choose the local gates to be orthogonal, these circuits generally anticoncentrate in logarithmic depth \cite{sauliere_universality_2025,grevink_will_2025}, but relative-error state 2-designs require linear depth \cite{grevink_will_2025}.
In contrast, our result clearly holds for Clifford circuits.
For symplectic circuits, however, the universality of anticoncentration remains open.

Although our results do not straightforwardly extend to the unitary design case, we believe that this is a limitation of the proof technique, and that a direct relation between anticoncentration and relative error unitary 2-designs should be provable under at least the same conditions for which Theorem \ref{thm:main} can be proven. 
This is supported by the numerical findings in Ref.~\cite{belkin_absence_2025}.

Finally, it would be interesting to understand whether 
higher-order designs can be similarly reduced to a small number of universal properties.
There, our technique fails as it strongly relies on the non-negative representation of local RQCs in the local permutation basis, which holds for second moments only.

\section*{Acknowledgements}

We thank D. Belkin for fruitful discussions on the unitary design case and for sharing an earlier draft of their work \cite{belkin_absence_2025}. 
M.\,H.~acknowledges funding by the Deutsche Forschungsgemeinschaft (DFG, German Research Foundation) - 54759578 and by the Deutsche Forschungsgemeinschaft (DFG, German Research Foundation) under Germany’s Excellence Strategy – Cluster of Excellence Matter and Light for Quantum Computing (ML4Q) EXC 2004/1 – 390534769. J.\,Helsen acknowledges funding from the Dutch Research Council (NWO) through a Veni grant (grant No.VI.Veni.222.331) and the Quantum Software Consortium (NWO Zwaartekracht Grant No.024.003.037). The result on state designs was derived by the authors during the 2024 Random Quantum Circuits workshop in Amsterdam.

\bibliography{notes.bib}

\clearpage

\onecolumngrid
\pdfbookmark{Supplemental material}{Supplemental material}

\begin{center}
\textbf{\large \thetitle}\\[1em]
\textit{\large -- Supplemental material --}
\end{center}


\setcounter{equation}{0}
\setcounter{figure}{0}
\setcounter{table}{0}
\setcounter{section}{0}
\setcounter{theorem}{0}
\makeatletter
\renewcommand{\theequation}{T\arabic{equation}}
\renewcommand{\thefigure}{T\arabic{figure}}
\renewcommand{\thesection}{T\Roman{section}}
\renewcommand{\thetheorem}{T\arabic{theorem}}

\section{Expansions in the local permutation basis}
\label{sec:SM:basis-expansions}

Using Schur's lemma and the projectors onto the globally symmetric or antisymmetric subspaces $P_{\Sym/\Alt} = \frac12 (\one + F_1)$ with dimensions $D_{\Sym/\Alt} = q^n (q^n\pm 1)/2$, respectively, we find:
\begin{align}
 \MO_\haar(F_x)
 &=
 \frac{\tr(\Psym F_x)}{\Dsym} \Psym + \frac{\tr(\Palt F_x)}{\Dalt} \Palt \label{eq:Haar-twirl} \\ 
 &=
 \left( \frac{ \tr(F_x) + \tr(F_1 F_x) }{4\Dsym} + \frac{ \tr(F_x) - \tr(F_1 F_x) }{4\Dalt} \right) F_0
 +
 \left( \frac{ \tr(F_x) + \tr(F_1 F_x) }{4\Dsym} - \frac{ \tr(F_x) - \tr(F_1 F_x) }{4\Dalt} \right) F_1 \\ 
 &=
 \frac{ q^{2n-|x|} - q^{|x|} }{q^{2n}-1} F_0 + \frac{ q^{n+|x|} - q^{n-|x|} }{q^{2n}-1} F_1 \\
 &=:
 h_{0,x}  F_0 + h_{1,x}  F_1 
\end{align}
In particular,
\begin{equation}
 h_{0,x}  + h_{1,x}  
 =
 \frac{ q^{2n-|x|} - q^{|x|} + q^{n+|x|} - q^{n-|x|} }{q^{2n}-1}
 =
 \frac{ (q^n - 1) (q^{n-|x|} + q^{|x|} ) }{q^{2n}-1}
 =
 \frac{q^{n-|x|} + q^{|x|}}{q^{n}+1} \,.
\end{equation}
We can write the dual basis explicitly by noting that the local permutation basis factorizes and by using the single-qudit Weingarten matrix:
\begin{align}
 \hat F_x &= \bigotimes_{i=1}^n \hat F_{x_i} \,, &
 \hat F_{x_i} &= \sum_{y_i} w_{x_i,y_i} F_{y_i} \,, &
 w &:= \frac{1}{q^2-1} \begin{pmatrix} 1 & -1/q \\ -1/q & 1 \end{pmatrix} \,.
\end{align}
In particular, $\hat F_0 = \frac{1}{q^2-1}( \one - \frac1q F )$ and $\hat F_1 = \frac{1}{q^2-1}( - \frac1q \one  + F )$.

\section{Some identities in the local projector basis}
\label{sec:some_identities_in_the_local_irrep_basis}

We can decompose any psd operators $A,B$ which are invariant under local unitaries (LU) by Schur's lemma as 
\begin{align}
  A &= \sum_{a\in\F_2^n} \frac{A_a}{D_a} P_a \,, &
  B &= \sum_{b\in\F_2^n} \frac{B_b}{D_b} P_b \,.
\end{align}
Here, $A_a = \tr(A P_a) \geq 0$ and $B_b = \tr(B P_b) \geq 0$.
Then, assuming that
\begin{equation}
| \tr( P_a \MO_\nu(P_b) )   - \tr( P_a \MO_\haar(P_b) )  |
\leq
\delta \tr( P_a \MO_\haar(P_b) ) \,,
\end{equation}
we find using triangle inequality and positivity of the coefficients:
\begin{equation}
| \tr(A \MO_\nu(B) )   - \tr(A \MO_\haar(B) )  |
\leq 
\sum_{a,b}  \frac{A_a B_b}{D_a D_b} | \tr( P_a \MO_\nu(P_b) )   - \tr( P_a \MO_\haar(P_b) )  |
\leq
\delta \tr(A \MO_\haar(B) ) \,.
\end{equation}

As in Eq.~\eqref{eq:Haar-twirl}, we can compute the Haar moments for $A=P_a$ and $B=P_b$ and find that the projectors have to have support in the same global irrep, meaning that $|a|$ and $|b|$ have to be both even or both odd.
Then, 
\begin{equation}
  \tr( P_a \MO_\haar(P_b)) = D_a D_b \times
  \begin{cases} 
   \Dsym^{-1} & \text{if } |a|,|b| \text{ even} \\ 
   \Dalt^{-1} & \text{if } |a|,|b| \text{ odd} \\ 
   0 & \text{else} 
  \end{cases} \,.
\end{equation}
We note that the LU-invariance of $\nu$ also implies that $\tr(P_a\MO_\nu(P_b)) = 0$ whenever the parity of $|a|$ and $|b|$ are not equal.
To see this, note that by the definition of $\MO_\nu$, we have $\MO_\nu(F_1 B) = F_1 \MO_\nu(B)$.
Recall that $F_1$ is the global flip, thus $\tr(P_a \MO_\nu(P_b)) = \tr(P_a \MO_\nu(P_{\Sym/\Alt} P_b)) = \tr(P_a P_{\Sym/\Alt} \MO_\nu( P_b))$ where $\Sym/\Alt$ is chosen according to the parity of $b$.

Note that the $\{P_a\}$ basis is orthogonal and $\hat P_a = P_a/D_a$ is its dual basis.
This basis is exactly the Fourier transform of the local permutation basis $\{F_x\}$:
\begin{align}
    P_a &= \bigotimes_{i=1}^n \left( \one + (-1)^{a_i} F \right) = \sum_{x\in\F_2^n} (-1)^{a\cdot x} F_x \,, &
    \hat P_b &= \frac{1}{D_b} P_b = \sum_{x\in\F_2^n} (-1)^{a\cdot y} \hat F_y \,.
\end{align}
Hence the matrix representation of of $\MO_\nu$ in the local projector basis, $\tilde m_{a,b} = \tr(\hat P_a \MO_\nu(P_b)) = \tr(P_a \MO_\nu(P_b))/D_a$, is just the Fourier transform of its representation in permutation basis.
The latter can be understood as a non-negative function on $\F_2^n \times \F_2^n$.
We can thus invoke Bochner's theorem to conclude that the matrix $\tilde A^{a,b}_{c,d} := \tilde m_{a+c,b+d}$ is psd.
In particular, we have the non-negativity of the principal minor
\begin{equation}
    0 \leq 
    \begin{vmatrix}
        \tilde m_{0,0} & \tilde m_{a+c,b+d} \\
        \tilde m_{a+c,b+d} & \tilde m_{0,0}
    \end{vmatrix}
    =
    \tilde m_{0,0}^2 - \tilde m_{a+c,b+d}^2 \,,
\end{equation}
thus $\tilde m_{0,0} \geq \tilde m_{a,b}$ for all $a,b$. 
We can then write Eq.~\eqref{eq:relative-error} of the main text as 
\begin{equation}
    \varepsilon = \frac{q^n}{2} \max_a \tilde m_{a,a}  \frac{q^n + (-1)^{|a|}}{2 D_a} - 1 \,.
\end{equation}
As we have shown above the term 
$\tilde m_{a,a}$ is maximized by $a=0$, while the maximum of the other, at least for large $n$, is given by $a=1$.

\end{document}